\documentclass[a4paper,fleqn,11pt]{article}
\usepackage[margin=1in]{geometry}
\usepackage{float}
\usepackage[utf8]{inputenc}
\usepackage[T1]{fontenc}
\usepackage{amsmath}
\usepackage{amssymb}
\usepackage{graphicx}
\usepackage{booktabs}
\usepackage{array}
\usepackage{multirow}
\usepackage[numbers,sort&compress]{natbib}
\usepackage[protrusion=true,expansion=false]{microtype}
\usepackage{placeins}
\usepackage{longtable}
\usepackage{authblk}
\usepackage{tabularx}
\newcolumntype{Y}{>{\raggedright\arraybackslash}X}
\newcommand{\incar}[1]{\texttt{#1}}
\usepackage[hidelinks]{hyperref}
\title{Fine-Tuning Small Language Models for Reliable VASP INCAR Generation}

\author[1]{Xinyue Zhang}
\author[2]{Jixiang Li}
\author[1,3,4]{Bin Shao\thanks{Corresponding author. Email: \texttt{bshao@nankai.edu.cn}}}
\author[5]{Baishun Yang}
\author[1,3,4]{Zhiyang Liu\thanks{Corresponding author. Email: \texttt{liuzhiyang@nankai.edu.cn}}}
\author[1,4,6,7]{Weichao Wang\thanks{Corresponding author. Email: \texttt{weichaowang@nankai.edu.cn}}}

\affil[1]{College of Electronic Information and Optical Engineering, Nankai University, Tianjin 300350, China}
\affil[2]{School of Physics, Nankai University, Tianjin 300071, China}
\affil[3]{Tianjin Key Laboratory of Optoelectronic Sensor and Sensing Network Technology, Nankai University, Tianjin 300350, China}
\affil[4]{Shenzhen Research Institute of Nankai University, Shenzhen 518057, China}
\affil[5]{CIC nanoGUNE BRTA, Tolosa Hiribidea 76, 20018 San Sebasti\'an, Spain}
\affil[6]{College of Energy and Environment Science, Yunnan Normal University, Kunming 650500, China}
\affil[7]{Southwest United Graduate School, Kunming 650092, China}

\date{}

\begin{document}

\maketitle

\begin{abstract}
Language models can prepare VASP INCAR files from natural-language requests, but so far only large proprietary cloud models come close to handling the tightly coupled, physics-sensitive settings reliably, a dependence that fits poorly with local, high-throughput materials workflows where privacy, cost, and offline deployment matter. We show that a small language model (SLM) can close this gap. The SLM is fine-tuned on reference VASP calculations and paired with VASPGuard, a deterministic post-processor that checks syntax, workflow, and material-dependent constraints; we call the combined model INCAR-SLM. On INCARBench, a benchmark for VASP INCAR generation, INCAR-SLM built on Qwen3-4B outperforms every general-purpose LLM evaluated, exceeding GPT-5.4 by 15.55 points on the 100-point INCAR Score. Most of this gain comes from fine-tuning, with VASPGuard correcting the errors that remain. We further find that model size matters less than expected: once fine-tuning and post-processing are applied, performance saturates at a few billion parameters, and Qwen3-4B outperforms larger models in the same family.
\end{abstract}


\vspace{1em}

\section{Introduction}

Large language models (LLMs) are changing how researchers interact with computational software and workflows \cite{lei2024materials}. Beyond question answering, they now support natural-language interfaces for scientific tasks \cite{prince2024toolaugmented}: a researcher states an objective, and the model retrieves knowledge, writes code, or prepares simulation inputs. In condensed-matter physics, materials science, and chemistry, this shift has driven work on predictive modelling, materials discovery, and foundation models \cite{jablonka2024predictive,jiang2025nlp,pyzerknapp2025foundation}, built on structural databases such as the Materials Project \cite{jain2013materialsproject,ong2015mpapi}, high-throughput calculations \cite{merchant2023scaling}, workflow libraries such as pymatgen \cite{ong2013pymatgen} and atomate2 \cite{ganose2025atomate2}, and LLM-based scientific agents \cite{jablonka2023hackathon,caldasramos2025review}. A natural target for this assistance is the input configuration of density functional theory (DFT) calculations \cite{hohenberg1964inhomogeneous,kohn1965selfconsistent}. In the Vienna Ab initio Simulation Package (VASP) \cite{kresse1994germanium,kresse1996iterative,kresse1996efficiency}, the INCAR file \cite{vaspwiki} encodes a calculation request as a set of tightly coupled, physics-dependent settings, so a file that runs is not necessarily scientifically correct. We previously quantified this difficulty with INCARBench \cite{incarbench2026}, a benchmark for INCAR generation and repair, and found that only large general-purpose models accessed as commercial online services come close to handling these couplings reliably.

For routine first-principles work, however, cloud dependence is a poor fit. Production calculations run on local, often offline clusters; structures and results must stay in-house; and batch generation of input files makes token-metered API costs, rate limits, and network latency add up quickly: each request carries the full structure and instructions, so charges grow with every material screened. What is needed instead is a model that is reliable and \emph{locally deployable}: small enough to run on local hardware, inexpensive to call at high throughput, and fully under the user's control. Providing such a model, rather than relying on an external service, is the problem we address.

Here we meet this need with a small language model (SLM), a model compact enough to run on local hardware. We fine-tune open SLMs on the INCAR Training Set to learn common calculation types. Each draft is then passed through VASPGuard, a deterministic, Materials Project-aligned post-processor that checks rule-governed INCAR constraints and corrects them. We refer to the resulting model as INCAR-SLM. Evaluated on INCARBench \cite{incarbench2026} under the same scoring protocol, INCAR-SLM built on Qwen3-4B reaches an INCAR Score of 89.88, ahead of every general-purpose LLM evaluated here, while remaining compact enough to run on a single GPU.
\section{Methods}

\FloatBarrier
\subsection{Workflow Overview}

\begin{figure}[htbp]
  \centering
  \includegraphics[width=\linewidth]{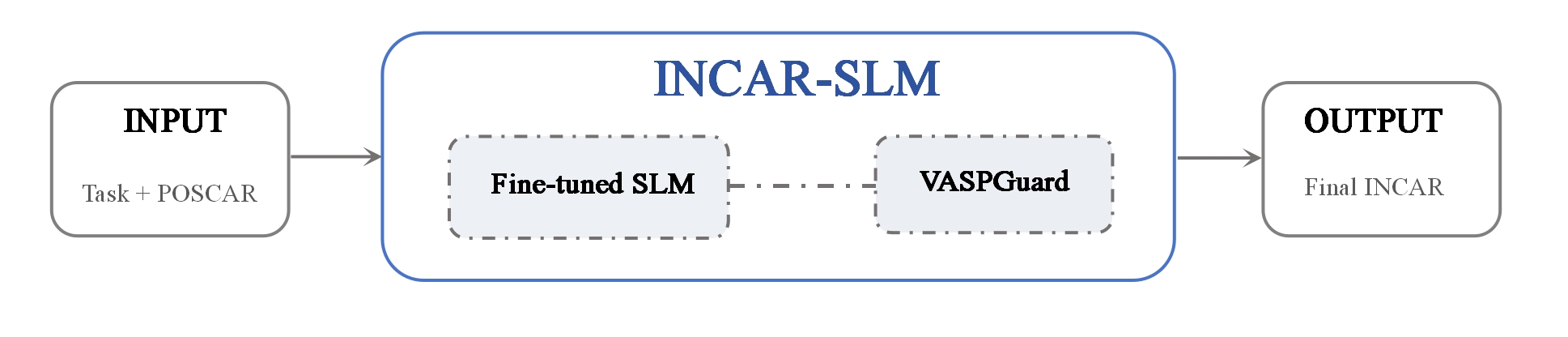}
  \caption{Overview of the INCAR-SLM workflow. Given a calculation task and a POSCAR structure, a fine-tuned SLM produces an INCAR draft, which VASPGuard checks and corrects to produce the final INCAR.}
  \label{fig:overview}
\end{figure}

We treat VASP INCAR generation as a constrained input-generation task rather than free-form text generation. As shown in Figure~\ref{fig:overview}, INCAR-SLM takes a calculation task and a POSCAR structure as input and returns a final INCAR as output. Internally, a fine-tuned SLM first generates an INCAR draft for the requested calculation type, such as geometry relaxation, a static self-consistent calculation, or a non-self-consistent band-structure or density-of-states (DOS) calculation. The draft is then checked by VASPGuard, a post-processing step that parses it into valid INCAR entries, verifies that workflow controls match the requested task, and applies material-dependent checks for settings such as DFT+$U$, magnetism, smearing, and vdW treatment. The result is a checked INCAR rather than an unfiltered model response.

The design separates generation from post-processing. INCAR tags are defined by the VASP manual, but the required settings are not fully determined by the POSCAR alone---they also depend on the calculation task and the material. Fine-tuning teaches the SLM to generate a suitable draft from the request and the structure; VASPGuard then applies explicit rules to correct what can be determined reliably, without a second model call. We describe each component in turn, starting with the data used for fine-tuning.
\FloatBarrier

\subsection{The INCAR Training Set}

The INCAR Training Set was built from reference VASP calculation records, normalized into a common input--output format. The underlying calculations follow standard plane-wave DFT practice, using projector augmented-wave pseudopotentials \cite{blochl1994paw,kresse1999ultrasoft} and the PBE exchange--correlation functional \cite{perdew1996pbe}, consistent with Materials Project conventions \cite{jain2013materialsproject}. Current language models remain uneven on materials-science knowledge \cite{zaki2024mascqa,bajan2025expertise}, and prior work on chemistry and materials adaptation shows that task-specific tuning is usually needed for reliable scientific generation \cite{vanherck2025assessment,choi2024materialslanguage}. The training data were therefore designed to cover both routine INCAR templates and the failure-prone settings that depend on task or material context.

The set covers four task types: static self-consistent calculations, geometry relaxations, and non-self-consistent band-structure and DOS calculations, following standard VASP distinctions in charge-density inheritance, ionic-motion control, and post-processing behavior \cite{vaspwiki}. Within each type, sampling favors cases where the correct INCAR cannot be determined from the task label alone: transition-metal compounds that need element-aligned DFT+$U$ arrays \cite{dudarev1998lsdau}, magnetic systems that need spin initialization, layered systems that need dispersion corrections \cite{grimme2010dftd} or nonlocal van der Waals (vdW) treatment \cite{klimes2011vdw}, and band or DOS calculations that need the correct reciprocal-space controls \cite{monkhorst1976special}. These cases target common LLM failure modes, where syntactically plausible INCAR files still carry incorrect workflow states, wrong element-wise arrays, missing MAGMOM, or incompatible symmetry settings.

Each case was organized as a standardized input--output example, linking the material structure, calculation type, Materials Project task identifier \cite{jain2013materialsproject}, and reference INCAR. The target output is a plain INCAR key--value list following the conventions used by pymatgen \cite{ong2013pymatgen}, keeping the training target close to practical VASP usage while retaining enough provenance for consistency checks against established materials-data practices \cite{ong2015mpapi}. The final set contains 4,453 cases: 1,091 static self-consistent, 1,130 geometry relaxation, 1,119 band-structure non-self-consistent, and 1,113 DOS non-self-consistent. This task balance is combined with enriched coverage of DFT+$U$, magnetism, symmetry control, non-self-consistent post-processing, and layered-material settings. 

\subsection{Domain Fine-Tuning}

Domain fine-tuning fits the model to a set of reference VASP calculations. Each training case pairs the physical specification of a calculation---the user's request, the calculation type, the material family, and the POSCAR structure---with the corresponding reference INCAR, itself represented as a sequence of tokens. For the $i$-th case, let $x_i$ denote the calculation specification and $y_i$ the reference INCAR. The model generates the INCAR one token at a time, so the probability it assigns to the complete file is
\begin{equation}
p_{\theta,\phi}(y_i \mid x_i) = \prod_{t=1}^{T_i} p_{\theta,\phi}(y_{i,t} \mid x_i, y_{i,<t}),
\label{eq:conditional}
\end{equation}
where $y_{i,t}$ is the $t$-th token of the reference INCAR and $T_i$ is the number of target tokens. During training, the model is shown the correct preceding tokens at each step; during inference, it instead conditions on its own previous output. Training minimizes the standard cross-entropy between the model's predictions and the reference tokens, so that higher probability is assigned to the correct INCAR settings; the calculation request and POSCAR provide context but are not themselves prediction targets. The complete loss function is given in Supplementary Information, Section~S3.

We use LoRA (low-rank adaptation) \cite{hu2021lora} for efficient optimization: the pre-trained weights $\theta$ remain fixed, and only a small set of low-rank correction weights $\phi$ is trained, substantially reducing memory and compute requirements while preserving the general knowledge in the pre-trained model. The rank, scaling factor, and target layers are given in Supplementary Information, Section~S2.

Fine-tuning proceeds in two stages: the first establishes the standard INCAR format and the main differences among geometry relaxation, static self-consistent, band-structure, and DOS calculations; the second focuses on cases where several settings must remain physically consistent, including element-dependent DFT+$U$ parameters, magnetic initialization, symmetry control, charge-density reuse in non-self-consistent DOS and band calculations, vdW treatment of layered materials, and ionic-relaxation settings. The data composition and training schedule for both stages are given in Supplementary Information, Sections~S4 and~S5. Fine-tuning alone, however, does not guarantee that every task-critical setting is correct, which motivates the post-processing step described next.

\subsection{VASPGuard}

To turn a model draft into the final INCAR, VASPGuard applies three steps: parsing, workflow correction, and composition-dependent correction. It first reads the draft together with the user request and the POSCAR, and converts it into a key--value representation; malformed or unparseable values are treated as syntax errors rather than completed automatically, so VASPGuard only acts on drafts that already resemble a valid INCAR. Workflow rules then check whether the control tags match the requested calculation type---for example, enforcing ionic-relaxation controls for geometry optimization, or the fixed-charge-density convention used for DOS and band-structure calculations. Composition-dependent rules are applied last, aligning DFT+$U$ arrays and magnetic settings with the POSCAR species and atom counts when the necessary evidence is available. Throughout, the rules encode the Materials Project protocol conventions followed by the reference calculations, with the VASP documentation providing the physical justification. The complete rule set is given in Supplementary Information, Section~S7.

To illustrate this division of labor, Table~\ref{tab:guard-example} shows a NaCl geometry-relaxation case. The fine-tuned model set \texttt{IBRION=-1} and \texttt{NSW=0}, disabling ionic motion, so the requested relaxation could not proceed. VASPGuard changed these to \texttt{IBRION=2} and \texttt{NSW=99}, enabling ionic motion, and left the remaining, already-compatible settings unchanged. Additional cases covering DFT+$U$, magnetism, and DOS workflows are given in Supplementary Information, Section~S8.

\begin{table}[H]
  \centering
  \caption{Representative VASPGuard correction for a NaCl geometry-relaxation task. Only selected INCAR entries are shown; check marks identify corrected settings.}
  \label{tab:guard-example}
  \begin{tabular}{llll}
    \toprule
    \multicolumn{2}{c}{Fine-tuned SLM output} & \multicolumn{2}{c}{Output after VASPGuard} \\
    \midrule
    ENCUT = 520 & ISIF = 3 & ENCUT = 520 & ISIF = 3 \\
    PREC = Accurate & NELM = 200 & PREC = Accurate & NELM = 200 \\
    IBRION = -1 ($\times$) & NSW = 0 ($\times$) & IBRION = 2 ($\checkmark$) & NSW = 99 ($\checkmark$) \\
    \bottomrule
  \end{tabular}
\end{table}
\FloatBarrier

\subsection{Evaluation on INCARBench}

We evaluate INCAR-SLM on INCARBench \cite{incarbench2026}, a benchmark for VASP INCAR generation built from Materials Project calculation records \cite{jain2013materialsproject,ong2013pymatgen}. INCARBench scores each generated INCAR against a normalized reference using two components: Must match, which checks whether task-critical entries---relaxation controls, charge-inheritance settings for DOS or band workflows, DFT+$U$, MAGMOM, and symmetry-related parameters---are explicit and semantically correct, with missing keys, wrong task states, incorrect Boolean switches, or mismatched element-wise arrays scored as zero; and Policy match, which checks whether general numerical and physical settings, such as ENCUT, convergence criteria, smearing, aspherical corrections, mixing parameters, and DOS-grid controls, fall within case-specific tolerances. The final INCAR Score averages the two,
\begin{equation}
S_{\mathrm{INCAR}} = \frac{S_{\mathrm{must}} + S_{\mathrm{policy}}}{2},
\label{eq:score}
\end{equation}
where $S_{\mathrm{must}}$ and $S_{\mathrm{policy}}$ are the Must match and Policy match percentages. The full benchmark composition and scoring tolerances are given in Supplementary Information, Section~S6.

To keep the comparison fair, every model is evaluated under the same task definitions, prompt templates, parsing procedures, and scoring scripts. We test three families of openly available small models across a range of sizes---Qwen3 \cite{yang2025qwen3}, Llama-3 \cite{dubey2024llama3}, and Gemma-3 \cite{gemmateam2025gemma3}---under four conditions: the foundation model alone, the foundation model with VASPGuard, the fine-tuned model alone, and the fine-tuned model with VASPGuard. The last condition is INCAR-SLM. Comparing these four conditions separates the contribution of domain fine-tuning from that of post-processing.

\section{Results and Analysis}

\subsection{Overall Performance against LLMs}

\begin{figure}[htbp]
  \centering
  \includegraphics[width=0.92\linewidth]{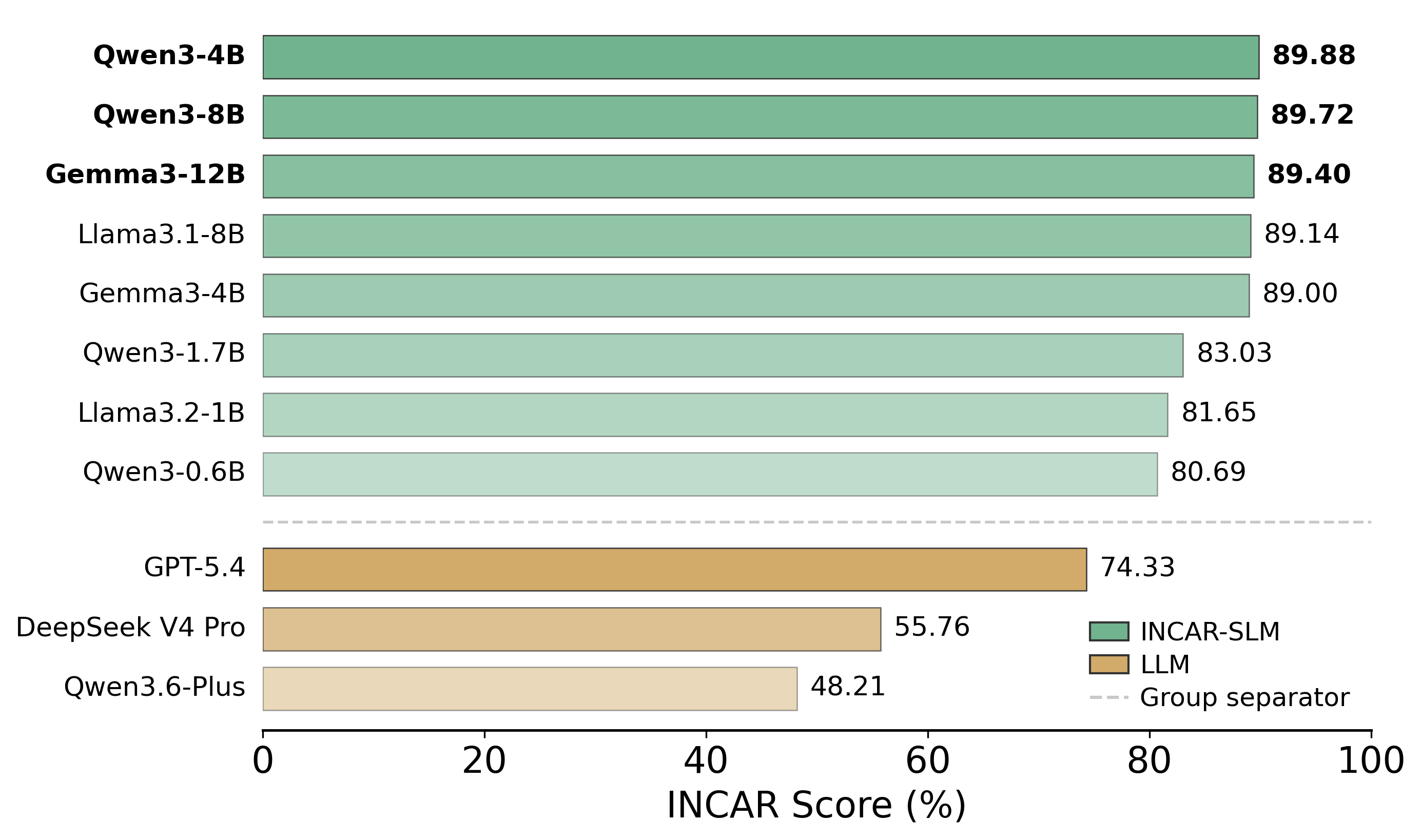}
  \caption{Ranking of final INCAR Scores. INCAR-SLM models and general-purpose LLMs are ranked by INCAR Score. The dashed line separates the local INCAR-SLM models from the LLMs.}
  \label{fig:ranking}
\end{figure}

We present the overall comparison between INCAR-SLM and the general-purpose LLMs on INCARBench in Figure~\ref{fig:ranking} and Table~\ref{tab:full-results}; model names are abbreviated here and throughout, with full identifiers given in Supplementary Information, Table~S2. INCAR-SLM outperforms every LLM evaluated: built on Qwen3-4B, it reaches the best score, 89.88, closely followed by Qwen3-8B (89.72), Gemma3-12B (89.40), Llama3.1-8B (89.14), and Gemma3-4B (89.00). Smaller base models also perform well: Qwen3-1.7B reaches 83.03, Llama3.2-1B reaches 81.65, and Qwen3-0.6B reaches 80.69. The general-purpose LLMs score lower across the board: 74.33 for GPT-5.4, 55.76 for DeepSeek V4 Pro, and 48.21 for Qwen3.6-Plus.

Built on Qwen3-4B, INCAR-SLM exceeds GPT-5.4 by 15.55 points while remaining small enough for local deployment. More importantly, the top-performing base models all cluster within a narrow range near 89--90 points.

{\small
\begin{longtable}{lrrrr}
  \caption{INCARBench results across model states. Must and Policy denote the Must match and Policy match percentages; Score is their average (Eq.~\ref{eq:score}). Parameter counts are listed where applicable. Model names are abbreviated; full model identifiers are given in Supplementary Information, Table~S2.}
  \label{tab:full-results}\\
  \toprule
  Model & Params & Must & Policy & Score \\
  \midrule
  \endfirsthead
  \multicolumn{5}{l}{{\footnotesize Table~\thetable\ (continued)}}\\
  \toprule
  Model & Params & Must & Policy & Score \\
  \midrule
  \endhead
  \midrule
  \multicolumn{5}{r}{{\footnotesize continued on next page}}\\
  \endfoot
  \bottomrule
  \endlastfoot
  \multicolumn{5}{l}{\itshape LLMs}\\*
  DeepSeek V4 Pro   & --   & 61.36 & 50.16 & 55.76 \\*
  GPT-5.4           & --   & 72.80 & 75.85 & 74.33 \\*
  Qwen3.6-Plus      & --   & 39.66 & 56.77 & 48.21 \\
  \midrule
  \multicolumn{5}{l}{\itshape Foundation}\\*
  Qwen3-0.6B        & 0.6B &  0.00 &  0.00 &  0.00 \\
  Qwen3-1.7B        & 1.7B &  5.96 & 15.25 & 10.60 \\
  Qwen3-4B          & 4B   & 41.52 & 53.80 & 47.66 \\
  Qwen3-8B          & 8B   & 51.69 & 45.65 & 48.67 \\
  Llama3.2-1B       & 1B   &  0.00 &  0.00 &  0.00 \\
  Llama3.2-3B       & 3B   &  6.01 &  0.70 &  3.36 \\
  Llama3.1-8B       & 8B   & 27.29 & 44.52 & 35.91 \\
  Gemma3-1B         & 1B   &  0.00 &  0.00 &  0.00 \\
  Gemma3-4B         & 4B   &  8.85 & 25.57 & 17.21 \\
  Gemma3-12B        & 12B  & 21.19 & 47.08 & 34.13 \\
  \midrule
  \multicolumn{5}{l}{\itshape Foundation + VASPGuard}\\*
  Qwen3-0.6B        & 0.6B &  0.00 &  0.00 &  0.00 \\
  Qwen3-1.7B        & 1.7B & 11.07 & 19.30 & 15.18 \\
  Qwen3-4B          & 4B   & 81.43 & 83.28 & 82.36 \\
  Qwen3-8B          & 8B   & 73.19 & 66.17 & 69.68 \\
  Llama3.2-1B       & 1B   &  0.00 &  0.00 &  0.00 \\
  Llama3.2-3B       & 3B   &  6.03 &  0.70 &  3.37 \\
  Llama3.1-8B       & 8B   & 63.26 & 72.24 & 67.75 \\
  Gemma3-1B         & 1B   &  0.00 &  0.00 &  0.00 \\
  Gemma3-4B         & 4B   & 34.32 & 44.78 & 39.55 \\
  Gemma3-12B        & 12B  & 77.71 & 85.92 & 81.82 \\
  \midrule
  \multicolumn{5}{l}{\itshape Domain Fine-Tuning}\\*
  Qwen3-0.6B        & 0.6B & 72.89 & 73.24 & 73.07 \\
  Qwen3-1.7B        & 1.7B & 63.78 & 72.29 & 68.03 \\
  Qwen3-4B          & 4B   & 80.80 & 70.95 & 75.87 \\
  Qwen3-8B          & 8B   & 82.35 & 78.20 & 80.27 \\
  Llama3.2-1B       & 1B   & 38.48 & 43.49 & 40.98 \\
  Llama3.2-3B       & 3B   & 39.39 & 61.64 & 50.51 \\
  Llama3.1-8B       & 8B   & 73.11 & 78.84 & 75.98 \\
  Gemma3-1B         & 1B   &  4.03 & 15.36 &  9.70 \\
  Gemma3-4B         & 4B   & 62.82 & 78.78 & 70.80 \\
  Gemma3-12B        & 12B  & 75.88 & 77.16 & 76.52 \\
  \midrule
  \multicolumn{5}{l}{\itshape Domain Fine-Tuning + VASPGuard (INCAR-SLM)}\\*
  Qwen3-0.6B        & 0.6B & 81.45 & 79.92 & 80.69 \\
  Qwen3-1.7B        & 1.7B & 82.56 & 83.50 & 83.03 \\
  \textbf{Qwen3-4B} & 4B   & \textbf{89.24} & \textbf{90.53} & \textbf{89.88} \\
  Qwen3-8B          & 8B   & 89.19 & 90.26 & 89.72 \\
  Llama3.2-1B       & 1B   & 82.56 & 80.74 & 81.65 \\
  Llama3.2-3B       & 3B   & 41.37 & 64.71 & 53.04 \\
  Llama3.1-8B       & 8B   & 87.94 & 90.33 & 89.14 \\
  Gemma3-1B         & 1B   & 11.73 & 21.08 & 16.40 \\
  Gemma3-4B         & 4B   & 88.73 & 89.27 & 89.00 \\
  Gemma3-12B        & 12B  & 89.03 & 89.76 & 89.40 \\
\end{longtable}
}

Table~\ref{tab:full-results} also shows that VASPGuard alone can close much of the gap to general-purpose LLMs: Qwen3-4B (82.36) and Gemma3-12B (81.82) with VASPGuard but no fine-tuning already exceed GPT-5.4 (74.33). This should not be read as evidence of reliability, however. Foundation models are prone to hallucinated or inconsistent settings on a structured task like INCAR generation, and post-processing alone cannot guarantee correctness across cases---it can only act on drafts that already contain enough valid information. The limitation is visible even at large scale: Gemma3-12B with VASPGuard (81.82) still falls short of the much smaller Qwen3-1.7B after fine-tuning and VASPGuard (83.03). Reliable INCAR generation therefore depends on fine-tuning rather than on parameter count or post-processing alone, which is what makes a small, locally deployable model such as INCAR-SLM practical.
\subsection{Contribution of Fine-Tuning and VASPGuard}

\begin{figure}[htbp]
  \centering
  \includegraphics[width=\linewidth]{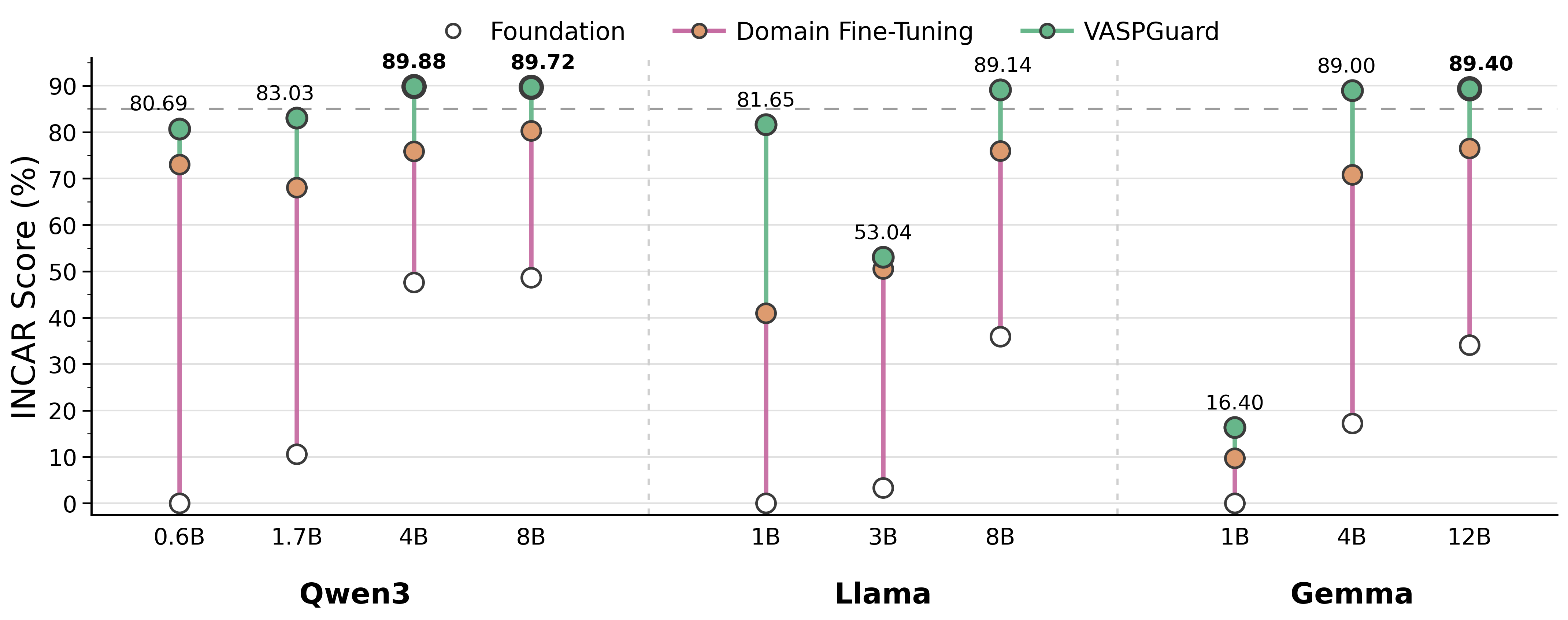}
  \caption{Contributions of domain fine-tuning and VASPGuard to the INCAR Score across model families and parameter sizes. Each vertical trajectory shows three states: the foundation model, the domain-fine-tuned model, and the final INCAR-SLM obtained after VASPGuard post-processing. The two connecting segments represent the successive gains from domain fine-tuning and post-processing. Final INCAR Scores are annotated above the green markers, with the three highest values shown in bold. The horizontal dashed line marks an INCAR Score of 85, and the vertical dashed lines separate the Qwen3, Llama, and Gemma families.}
  \label{fig:gain-trajectories}
\end{figure}

Having established the overall ranking, we next examine the contributions of domain fine-tuning and VASPGuard to the final INCAR Score. Figure~\ref{fig:gain-trajectories} organizes the Qwen3, Llama, and Gemma models by parameter size. Each vertical trajectory shows the foundation-model score, the score after domain fine-tuning, and the final INCAR-SLM score after VASPGuard. The first increase reflects the learning of task- and material-dependent INCAR settings, while the second reflects the correction of remaining rule-based errors by VASPGuard. All three states are evaluated using the same benchmark and scoring procedure, allowing the improvements to be compared directly across models.

Domain fine-tuning raises the INCAR Score for every SLM, and Figure~\ref{fig:gain-trajectories} shows the pattern clearly: the foundation-model markers cluster near the bottom of each family, particularly at the smallest sizes, where limited capacity and the absence of VASP-specific training keep several models at an INCAR Score of 0.00. Fine-tuning lifts every base model well above this floor---Qwen3-1.7B rises from 10.60 to 68.03, Llama3.2-1B from 0.00 to 40.98, and Gemma3-4B from 17.21 to 70.80, with comparable gains for Qwen3-4B, Llama3.1-8B, and Gemma3-12B (75.87, 75.98, and 76.52, respectively). VASPGuard then pushes most base models above the 85-point reference line marked in the figure, reaching 89.88 for Qwen3-4B, 89.72 for Qwen3-8B, 89.40 for Gemma3-12B, 89.14 for Llama3.1-8B, and 89.00 for Gemma3-4B. One base model breaks this pattern: Llama3.2-3B improves only from 50.51 to 53.04, well short of the 85-point line reached by every other mid-to-large model.

For Llama3.2-3B, 459 of the 500 drafts contained at least one invalid value---Boolean values for \texttt{IBRION}, multiple values for \texttt{ISIF}, non-integer values for \texttt{NELM}---so the evidence gate, which tolerates no invalid entries, passed only 41 cases and left VASPGuard able to clean up formatting but not reconstruct the calculation. The other nine models show no such failure, and all reach an INCAR Score above 80.

In addition, fine-tuning gains are larger for smaller models: Qwen3-0.6B and Qwen3-1.7B gain the most, while Qwen3-4B and Qwen3-8B gain less. Yet the smaller models still trail the larger ones in final score, since they start from a weaker general capacity. Choosing the best INCAR-SLM therefore means balancing fine-tuning gain against this remaining capacity gap.
\FloatBarrier
\subsection{Performance on Challenging Categories}

\begin{figure}[htbp]
  \centering
  \includegraphics[width=\linewidth]{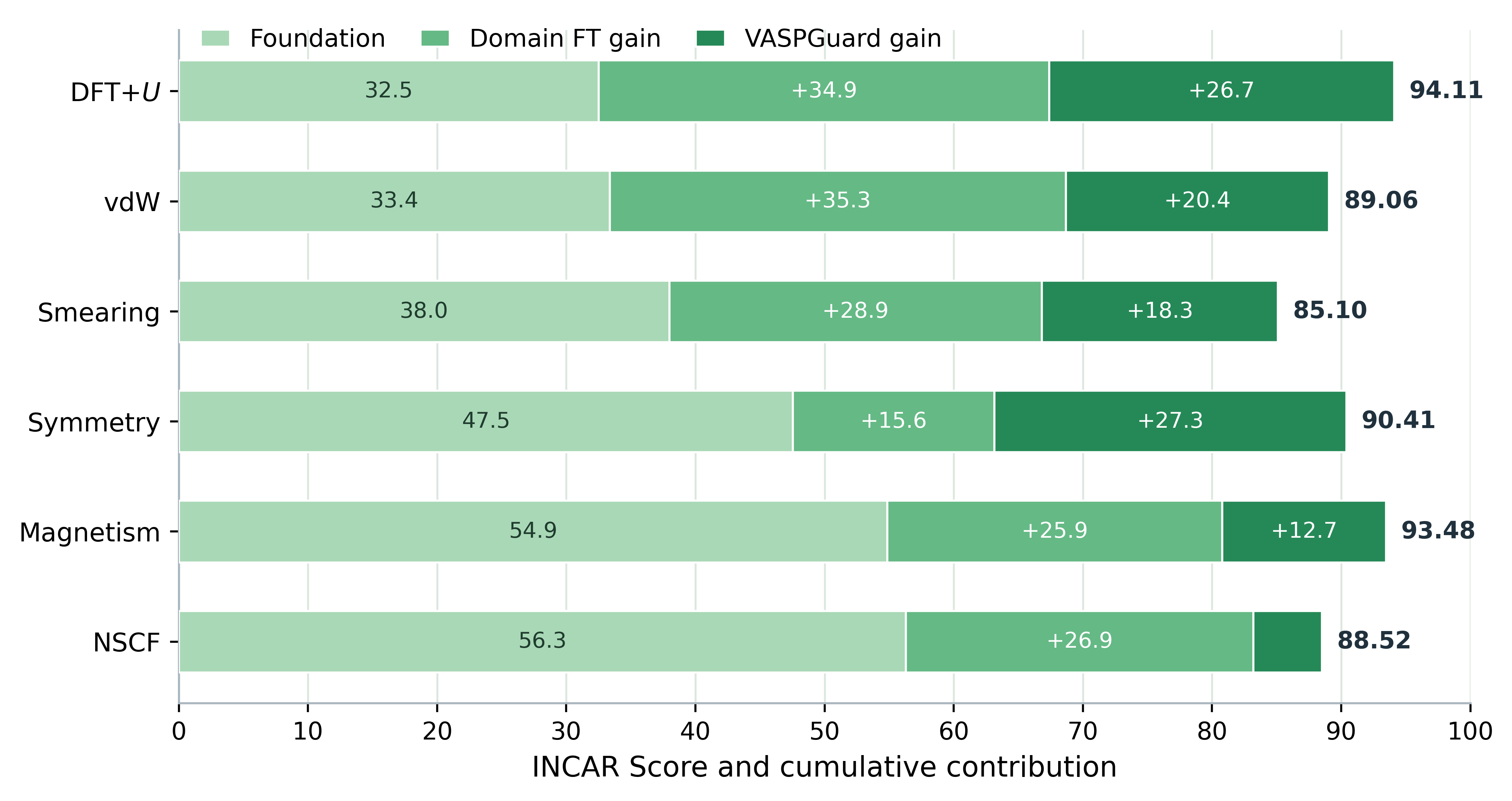}
  \caption{Decomposition of the INCAR Score across challenging categories for Qwen3-4B. Each bar starts with the foundation-model score (light green), adds the gain from fine-tuning (medium green), then adds the gain from VASPGuard (dark green); the bold number is the final score.}
  \label{fig:challenge}
\end{figure}

The overall score in Table~\ref{tab:full-results} averages over many tasks and does not show which settings are hardest to get right. In our previous work, errors from general-purpose models concentrated in a few material- and workflow-dependent categories rather than spreading evenly across INCAR tags. Figure~\ref{fig:challenge} breaks down the score for Qwen3-4B across six such categories: DFT+$U$, magnetism, van der Waals (vdW) treatment, non-self-consistent (NSCF) workflows, symmetry control, and smearing.

The foundation model struggles most with DFT+$U$ (32.50), vdW (33.36), and smearing (37.98), all of which need information beyond the calculation label itself. Fine-tuning raises every category, with the largest gains for vdW (+35.34), DFT+$U$ (+34.89), and smearing (+28.85).

VASPGuard adds a further, category-dependent gain, largest for symmetry (+27.27) and DFT+$U$ (+26.72), and smallest for NSCF (+5.31), since fine-tuning already resolves most of the workflow selection there. With both steps applied, the final scores range from 85.10 (smearing) to 94.11 (DFT+$U$). The pattern reflects a clear division of labor: fine-tuning learns which workflow and material context apply, while VASPGuard enforces constraints already explicit in the request, the POSCAR, or the INCAR schema. The total recovery is accordingly largest for DFT+$U$ (+61.61) and vdW (+55.70), and smallest for magnetism (+38.62) and NSCF (+32.22), whose foundation scores were already higher to begin with.
\FloatBarrier
\subsection{Dependence on Model Size}

\begin{figure}[htbp]
  \centering
  \includegraphics[width=\linewidth]{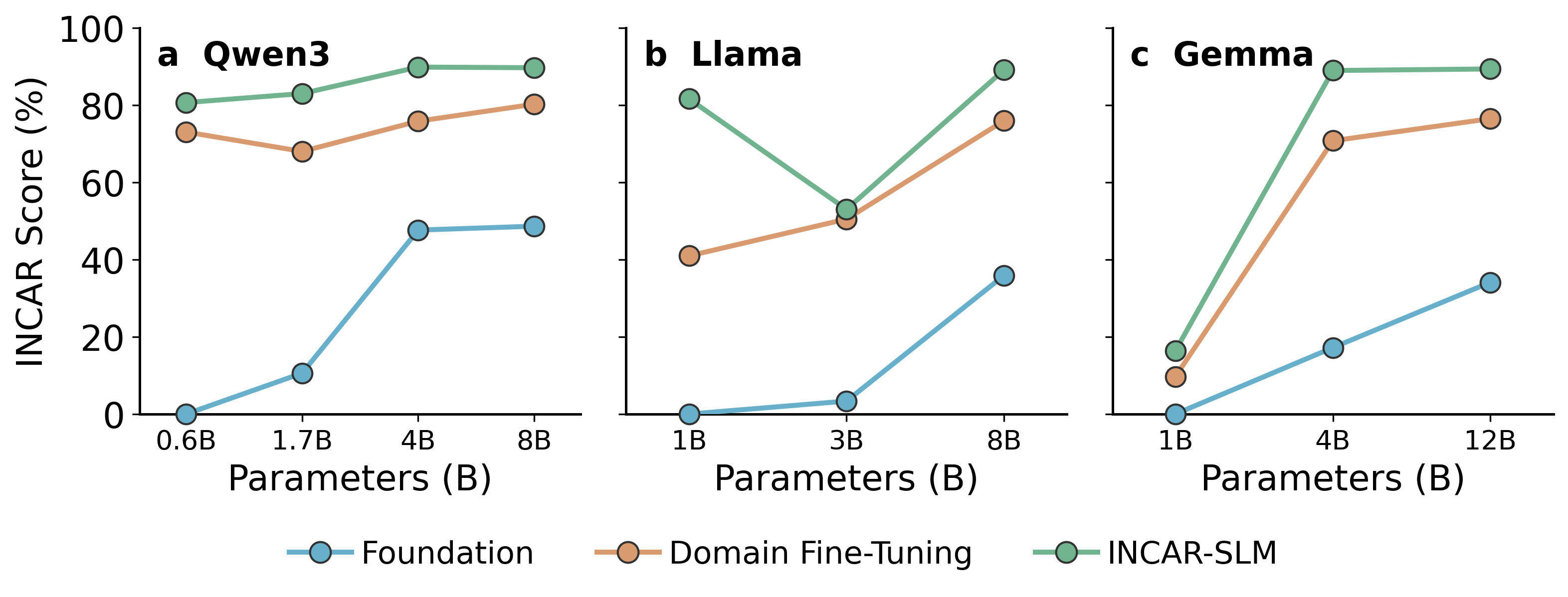}
  \caption{Model scaling trends under three states: foundation, domain fine-tuning, and the INCAR-SLM (domain fine-tuning with VASPGuard).}
  \label{fig:scaling}
\end{figure}

A natural question is whether the improvement is mainly due to increasing the model size. Here we find that scaling parameter count alone does not explain this improvement: the best foundation-model score reaches only 48.67 for Qwen3, 35.91 for Llama, and 34.13 for Gemma (Figure~\ref{fig:scaling}). Domain fine-tuning drives most of the gain, teaching each base model the INCAR settings associated with different VASP workflows and material conditions; VASPGuard then corrects the remaining rule-checkable errors. The combination works best for Qwen3 and Gemma, reaching 89.88 and 89.72 for Qwen3-4B and Qwen3-8B, and 89.00 and 89.40 for Gemma3-4B and Gemma3-12B.

The results point to saturation at a few billion parameters rather than continued scaling: Qwen3-4B slightly outperforms Qwen3-8B, and Gemma3-4B already approaches Gemma3-12B. Llama3.2-3B remains the exception discussed above, limited by draft quality rather than parameter count. Overall, Qwen3-4B with fine-tuning and VASPGuard gives the best result, showing that reliable INCAR generation depends more on domain fine-tuning and VASPGuard than on model size alone.
\FloatBarrier

\section{Discussion}

INCAR-SLM differs from established workflow tools such as pymatgen \cite{ong2013pymatgen}, Atomate \cite{mathew2017atomate}, and Atomate2 \cite{ganose2025atomate2} mainly in what the user must provide. These tools are efficient once the calculation protocol is already decided, but the user must still select the workflow and specify material-dependent settings---for example, setting up a DFT+$U$ relaxation for a magnetic transition-metal oxide requires choosing Hubbard parameters, magnetic initialization, symmetry treatment, smearing, and convergence criteria, which demands substantial knowledge of VASP conventions. INCAR-SLM instead takes a calculation request and a POSCAR, and returns a case-specific, checked INCAR (Table~\ref{tab:pymatgen-comparison}): the user describes the objective rather than enumerating INCAR tags, and the resulting file can still be used within existing workflow frameworks. This request-to-input role is related to recent work on local materials language models \cite{takahara2025materialsrag}.

\begin{table}[htbp]
  \centering
  \caption{Difference in user input between existing workflow tools and INCAR-SLM.}
  \label{tab:pymatgen-comparison}
  \small
  \begin{tabular}{p{0.29\linewidth} p{0.29\linewidth} p{0.30\linewidth}}
    \toprule
    Method & User input & Required knowledge \\
    \midrule
    Pymatgen / Atomate / Atomate2
      & Detailed protocol and structure
      & VASP settings and workflow details \\
    INCAR-SLM
      & Calculation request and POSCAR
      & Calculation objective \\
    \bottomrule
  \end{tabular}
\end{table}
\FloatBarrier

The two approaches suit different stages of a project. Once a protocol is fixed, pymatgen and Atomate2 generate INCAR files by rule alone, giving identical settings across cases and integrating with workflow managers for large-scale job submission---properties that matter most for high-throughput screening. INCAR-SLM instead targets the earlier stage, before a protocol is fixed, and can hand its output to these tools once one is settled on.

The same approach may extend to other formats with explicit syntax and checkable constraints, such as Quantum ESPRESSO \cite{giannozzi2017quantumespresso}, ABINIT \cite{gonze2009abinit}, LAMMPS \cite{thompson2022lammps}, and OpenFOAM \cite{weller1998openfoam}, though it is less suited to open-ended tasks without explicit protocol requirements.
\section{Conclusions}

We presented INCAR-SLM, a locally deployable model that combines domain fine-tuning with VASPGuard, a deterministic post-processor, to generate VASP INCAR files from a calculation request and a POSCAR structure. INCAR-SLM reaches an INCAR Score of 89.88, 15.55 points above GPT-5.4; fine-tuning provides most of this improvement, and VASPGuard corrects the workflow- and material-dependent errors that remain, particularly for DFT+$U$ and magnetic calculations. Model size alone does not explain this result---Qwen3-8B, with twice the parameters, scores slightly lower---so a small model, adapted to the task and paired with rule-based correction, can match or exceed a much larger proprietary one. INCAR-SLM thus offers a practical route to reliable, local INCAR generation, while the final scientific judgment remains with the user.

\section*{Data Availability}

The INCAR Training Set, the INCARBench evaluation split, the fine-tuning and generation code, and the VASPGuard source are available at \url{https://github.com/zxy-boop/VASPINCAR-Gen}. Code is released under the MIT License; the datasets are released under the Creative Commons Attribution 4.0 International License (CC BY 4.0), in accordance with the attribution terms of the Materials Project data from which they are derived. Further details are given in Supplementary Information, Section~S9.

\section*{Acknowledgements}

This work was supported by the Yunnan Provincial Science and Technology United Graduate School (202402AO370001), the National Natural Science Foundation of China (Grant Nos. 12474235 and 22576107), the Key Program of Tianjin Natural Science Foundation (Grant No. 24JCZDJC01230), and the Supercomputing Center of Nankai University (NKSC).

\bibliographystyle{unsrtnat}
\bibliography{references}

\clearpage
\setcounter{section}{0}
\setcounter{table}{0}
\setcounter{equation}{0}
\renewcommand{\thesection}{S\arabic{section}}
\renewcommand{\thetable}{S\arabic{table}}
\renewcommand{\theequation}{S\arabic{equation}}
\begin{center}
{\LARGE\bfseries Supplementary Information}\\[1.2em]
\end{center}

\section{Scope}

This document provides the information needed to reproduce the training and evaluation of INCAR-SLM. It reports the model and LoRA parameter counts, the training data and optimization settings, the common inference and scoring procedure, the main VASPGuard rules, and three representative examples. The complete case files, configurations, and evaluation code are provided with the data and code release.

\section{Models and trainable weights}

The original model weights were frozen. Only LoRA weights attached to the
attention and feed-forward projections were optimized. Table~\ref{tab:model-weights}
reports parameter counts obtained directly from the stored model and adapter
tensors. The exact counts can differ slightly from the rounded model names.

\begin{table}[H]
  \centering
  \caption{Base-model and trainable LoRA parameter counts. The last two columns
  give the learning rates used in the first and second training stages.}
  \label{tab:model-weights}
  \small
  \begin{tabular}{lrrrrr}
    \toprule
    Model & Base params. & LoRA params. & Trainable (\%) & Stage 1 LR & Stage 2 LR \\
    \midrule
    Qwen3-0.6B            & 0.752B & 20.19M  & 2.686 & $5.0\times10^{-5}$ & $1.5\times10^{-5}$ \\
    Qwen3-1.7B            & 2.032B & 34.87M  & 1.716 & $5.0\times10^{-5}$ & $1.5\times10^{-5}$ \\
    Qwen3-4B              & 4.022B & 66.06M  & 1.642 & $5.0\times10^{-5}$ & $1.5\times10^{-5}$ \\
    Qwen3-8B              & 8.191B & 87.29M  & 1.066 & $5.0\times10^{-5}$ & $1.5\times10^{-5}$ \\
    Llama-3.2-1B-Instruct & 1.236B & 22.54M  & 1.824 & $5.0\times10^{-5}$ & $1.5\times10^{-5}$ \\
    Llama-3.2-3B-Instruct & 3.213B & 48.63M  & 1.514 & $5.0\times10^{-5}$ & $1.5\times10^{-5}$ \\
    Llama-3.1-8B-Instruct & 8.030B & 83.89M  & 1.045 & $5.0\times10^{-5}$ & $1.5\times10^{-5}$ \\
    Gemma-3-1B-IT         & 1.000B & 26.09M  & 2.609 & $8.0\times10^{-5}$ & $2.0\times10^{-5}$ \\
    Gemma-3-4B-IT         & 4.300B & 59.60M  & 1.386 & $8.0\times10^{-5}$ & $2.0\times10^{-5}$ \\
    Gemma-3-12B-IT        & 12.187B & 130.94M & 1.074 & $5.0\times10^{-5}$ & $1.5\times10^{-5}$ \\
    \bottomrule
  \end{tabular}
\end{table}

For every model, LoRA was applied to \incar{q\_proj}, \incar{k\_proj},
\incar{v\_proj}, \incar{o\_proj}, \incar{gate\_proj}, \incar{up\_proj}, and
\incar{down\_proj}. The LoRA rank was 32, the scaling parameter was 64, and
the dropout probability was 0.05. No full-parameter fine-tuning or quantized
training was used.

Table~\ref{tab:name-mapping} lists the correspondence between the abbreviated model names used in the main text and figures, and the full model identifiers used for the base-model releases in Table~\ref{tab:model-weights}.

\begin{table}[H]
  \centering
  \caption{Correspondence between the abbreviated model names used in the main text and the full model identifiers.}
  \label{tab:name-mapping}
  \small
  \begin{tabular}{ll}
    \toprule
    Short name (main text) & Full model identifier \\
    \midrule
    Qwen3-0.6B   & Qwen3-0.6B \\
    Qwen3-1.7B   & Qwen3-1.7B \\
    Qwen3-4B     & Qwen3-4B \\
    Qwen3-8B     & Qwen3-8B \\
    Llama3.2-1B  & Llama-3.2-1B-Instruct \\
    Llama3.2-3B  & Llama-3.2-3B-Instruct \\
    Llama3.1-8B  & Llama-3.1-8B-Instruct \\
    Gemma3-1B    & Gemma-3-1B-IT \\
    Gemma3-4B    & Gemma-3-4B-IT \\
    Gemma3-12B   & Gemma-3-12B-IT \\
    \bottomrule
  \end{tabular}
\end{table}

\section{Domain fine-tuning: training objective}

Each training case pairs the physical specification of a calculation $x_i$ --- the user's request, calculation type, material family, and POSCAR structure --- with a reference INCAR $y_i$, represented as a sequence of tokens. The model assigns a probability to the complete INCAR by factorizing it into per-token conditional probabilities,
\begin{equation}
p_{\theta,\phi}(y_i \mid x_i) = \prod_{t=1}^{T_i} p_{\theta,\phi}(y_{i,t} \mid x_i, y_{i,<t}),
\label{eq:si-conditional}
\end{equation}
where $y_{i,t}$ is the $t$-th token of the reference INCAR, $y_{i,<t}$ denotes the preceding reference tokens, and $T_i$ is the number of target tokens. During training, the preceding reference tokens are supplied to the model (teacher forcing); during inference, the model instead conditions on the tokens it has already generated.

The model is trained by minimizing the average negative log-probability of the reference INCAR tokens,
\begin{equation}
\mathcal{L}_{\mathrm{SFT}}(\phi) =
- \frac{1}{\sum_{i=1}^{N} T_i}
\sum_{i=1}^{N}\sum_{t=1}^{T_i}
\log p_{\theta,\phi}(y_{i,t} \mid x_i, y_{i,<t}),
\label{eq:si-sft}
\end{equation}
where $N$ is the number of training cases, $\theta$ denotes the frozen base-model weights, and $\phi$ denotes the trainable LoRA weights described in Section~S2. A lower value of $\mathcal{L}_{\mathrm{SFT}}$ means the model assigns higher probability to the correct INCAR settings. Only the reference INCAR contributes to the loss; the calculation request and POSCAR supply context but are not themselves prediction targets, so the optimization teaches the model how changes in the requested calculation and material structure map onto changes in the required VASP parameters.
\section{Training data}

The model input contains a natural-language calculation request and a POSCAR
structure. The target is an INCAR written as plain \incar{KEY = VALUE} lines.
The data cover four workflows: static self-consistent calculations, geometry
relaxations, line-mode band calculations, and DOS non-self-consistent
calculations. The split sizes are given in Table~\ref{tab:data-splits}.

\begin{table}[H]
  \centering
  \caption{Training-data composition. Stage 1 provides balanced examples of the
  four workflows. Stage 2 contains the full task-specific training set.}
  \label{tab:data-splits}
  \small
  \begin{tabular}{lrrrrr}
    \toprule
    Split & Static SCF & Relaxation & Bands NSCF & DOS NSCF & Total \\
    \midrule
    Stage 1 train      & 300  & 300  & 300  & 300  & 1,200 \\
    Stage 1 validation & 40   & 40   & 40   & 40   & 160 \\
    Stage 1 test       & 40   & 40   & 40   & 40   & 160 \\
    Stage 2 train      & 1,091 & 1,130 & 1,119 & 1,113 & 4,453 \\
    Stage 2 validation & 100  & 138  & 125  & 136  & 499 \\
    Stage 2 test       & 119  & 143  & 148  & 119  & 529 \\
    \bottomrule
  \end{tabular}
\end{table}

The release also contains 4,682 source-case directories. Each directory
includes the POSCAR, the source INCAR record, the normalized reference INCAR,
and provenance metadata. Benchmark structures were excluded during dataset
construction by a normalized POSCAR fingerprint. The final data validation
reported no benchmark-structure overlap and no malformed POSCAR, invalid INCAR
array length, or conflicting target for an identical prompt.

\section{Training and inference settings}

Training was performed in two stages. Stage 1 introduced the INCAR syntax and
the four calculation workflows. Stage 2 continued from the Stage 1 adapter and
used the complete task-specific data. Table~\ref{tab:training-settings}
summarizes settings shared by all models.

\begin{table}[H]
  \centering
  \caption{Shared training settings.}
  \label{tab:training-settings}
  \small
  \begin{tabular}{ll}
    \toprule
    Setting & Value \\
    \midrule
    Sequence length & 12,800 tokens \\
    Per-device batch size & 1 \\
    Gradient accumulation & 32 steps \\
    Effective batch size & 32 samples \\
    Stage 1 duration & 0.5 epoch \\
    Stage 2 duration & 2.0 epochs \\
    Optimizer & AdamW \\
    Learning-rate schedule & Cosine \\
    Warm-up ratio & 0.03 \\
    Maximum gradient norm & 1.0 \\
    Numerical precision & bfloat16 \\
    Memory control & Gradient checkpointing \\
    Stage 2 model selection & Lowest validation loss \\
    Random seed & 42 \\
    \bottomrule
  \end{tabular}
\end{table}

Each training run used one NVIDIA A800 GPU with 80~GB memory. The software
environment used PyTorch 2.6.0, Transformers 4.52.4, PEFT 0.15.2,
LLaMA-Factory 0.9.3, Datasets 3.6.0, and Accelerate 1.7.0.

For evaluation, every model received the same scientific request and POSCAR.
Generation was deterministic, with greedy decoding and a maximum of 768 new
tokens. Qwen3 reasoning output was disabled. The generated text was reduced to
recognized INCAR assignments before scoring; the unfiltered response was also
retained for audit. An empty extracted INCAR therefore means that the model
returned no valid INCAR setting, not necessarily that it returned no text.

\section{INCARBench evaluation}

INCARBench contains 500 cases: 107 static SCF, 137 geometry-relaxation,
127 line-mode band, and 129 DOS NSCF cases. For each case, Must match measures
agreement on task-defining settings, while Policy match evaluates numerical or
protocol settings using the tolerances specified by that case. The reported
score is

\begin{equation}
  \mathrm{INCAR\ Score} =
  \frac{\mathrm{Must\ match}+\mathrm{Policy\ match}}{2}.
\end{equation}

Optional tags, extra tags, runnable checks, and KPOINTS-policy checks were
retained as diagnostics but were not included in this score. All models were
evaluated on the same 500 cases with the same scorer. 

\section{VASPGuard rules and limits}

VASPGuard is applied only after model generation. It reads the generated
response and POSCAR. Full
workflow corrections are enabled only when the draft contains enough valid INCAR and workflow settings. Drafts with malformed values receive cleanup only;
VASPGuard does not reconstruct a calculation from an empty draft.

\begin{table}[H]
  \centering
  \caption{Main workflow settings applied by VASPGuard after a draft passes the
  evidence gate. Only the task-defining settings are listed.}
  \label{tab:guard-rules}
  \small
  \begin{tabularx}{\textwidth}{lY}
    \toprule
    Workflow & Main post-processing rules \\
    \midrule
    Geometry relaxation &
    \incar{IBRION=2}, \incar{NSW=99}, \incar{ISIF=3},
    \incar{EDIFFG=-0.02}; enforce suitable electronic convergence and smearing. \\
    Static SCF &
    \incar{NSW=0}, \incar{LCHARG=.TRUE.}, \incar{LORBIT=11},
    \incar{LREAL=.FALSE.}; enforce static-SCF convergence and smearing. \\
    Line-mode bands &
    \incar{IBRION=-1}, \incar{NSW=0}, \incar{ICHARG=11},
    \incar{ISYM=0}, \incar{ISMEAR=0}; do not write a new charge density. \\
    DOS NSCF &
    \incar{IBRION=-1}, \incar{NSW=0}, \incar{ICHARG=11},
    \incar{ISYM=2}, \incar{NEDOS=2001}, \incar{ISMEAR=-5}. \\
    \bottomrule
  \end{tabularx}
\end{table}

For supported correlated oxides and fluorides, VASPGuard writes
species-resolved Dudarev arrays and checks their length against the POSCAR
species list. The implemented $U$ values are 3.25, 3.7, 3.9, 5.3, 3.32, 6.2,
4.38, and 6.2~eV for V, Cr, Mn, Fe, Co, Ni, Mo, and W, respectively. Magnetic
initialization is added only when supported by the draft or request, and the
\incar{MAGMOM} length is checked against the POSCAR atom count. VASPGuard does
not automatically add or change spin--orbit coupling or \incar{IVDW}; these
choices are not uniquely determined by structure alone.

\section{Representative examples}

The following examples use three model families and three calculation tasks.
Only settings needed to show the main changes are listed; the scores are the
case-level values from the same scorer used for the full benchmark.

\subsection{Qwen3-4B: static SCF calculation for W}

The input requested a static ground-state calculation for elemental W and
included its POSCAR. The foundation model produced a generic draft with
\incar{IBRION=0} and no explicit \incar{NSW}. Fine-tuning supplied the static
workflow, while VASPGuard corrected the remaining convergence and smearing
settings.

\begin{table}[H]
  \centering
  \caption{Representative Qwen3-4B result for
  \incar{W\_Static\_SCF}.}
  \label{tab:case-qwen}
  \small
  \begin{tabularx}{\textwidth}{lrrrY}
    \toprule
    State & Must & Policy & Score & Representative settings \\
    \midrule
    Foundation & 0.00 & 50.00 & 25.00 &
    \incar{ENCUT=500}, \incar{IBRION=0}, \incar{ISMEAR=0} \\
    Domain FT & 75.00 & 50.00 & 62.50 &
    \incar{ENCUT=520}, \incar{IBRION=-1}, \incar{NSW=0},
    \incar{ISMEAR=0} \\
    Domain FT + VASPGuard & 75.00 & 100.00 & 87.50 &
    \incar{EDIFF=1e-5}, \incar{NSW=0}, \incar{ISMEAR=-5},
    \incar{SIGMA=0.05}, \incar{LCHARG=.TRUE.} \\
    \bottomrule
  \end{tabularx}
\end{table}

\subsection{Llama-3.2-1B-Instruct: geometry relaxation of
\texorpdfstring{WSe$_2$}{WSe2}}

The foundation model returned text, but none of it was a valid INCAR setting;
the extracted INCAR was therefore empty and received a score of zero. After
fine-tuning, the model generated the essential ionic controls. VASPGuard then
corrected the force threshold, smearing, and magnetic-array length.

\begin{table}[H]
  \centering
  \caption{Representative Llama-3.2-1B-Instruct result for
  \incar{WSe2\_Geometry\_Relax}.}
  \label{tab:case-llama}
  \small
  \begin{tabularx}{\textwidth}{lrrrY}
    \toprule
    State & Must & Policy & Score & Representative settings \\
    \midrule
    Foundation & 0.00 & 0.00 & 0.00 &
    No valid INCAR assignment was extracted from the raw response. \\
    Domain FT & 100.00 & 60.00 & 80.00 &
    \incar{IBRION=2}, \incar{NSW=99}, \incar{ISIF=3},
    \incar{EDIFFG=0.001} \\
    Domain FT + VASPGuard & 100.00 & 100.00 & 100.00 &
    \incar{IBRION=2}, \incar{NSW=99}, \incar{ISIF=3},
    \incar{EDIFFG=-0.02}, \incar{ISMEAR=-5} \\
    \bottomrule
  \end{tabularx}
\end{table}

\subsection{Gemma-3-4B-IT: DOS calculation for
\texorpdfstring{Cr$_2$O$_3$}{Cr2O3}}

The input specified a DOS calculation using a previously converged charge
density. Fine-tuning recovered most of the DOS workflow, but used
\incar{ICHARG=1} and omitted DFT+$U$. VASPGuard changed the charge-density mode,
added species-resolved Hubbard settings, and removed \incar{NBANDS}, which is
not inferred for this DOS workflow without information from a preceding run.

\begin{table}[H]
  \centering
  \caption{Representative Gemma-3-4B-IT result for
  \incar{Cr2O3\_DOS\_NSCF}.}
  \label{tab:case-gemma}
  \small
  \begin{tabularx}{\textwidth}{lrrrY}
    \toprule
    State & Must & Policy & Score & Representative settings \\
    \midrule
    Foundation & 0.00 & 14.29 & 7.14 &
    \incar{ENCUT=400}, \incar{IBRION=2}, \incar{NSW=100} \\
    Domain FT & 25.00 & 85.71 & 55.35 &
    \incar{IBRION=-1}, \incar{NSW=0}, \incar{ICHARG=1},
    \incar{NEDOS=2001}, \incar{NBANDS=77} \\
    Domain FT + VASPGuard & 100.00 & 100.00 & 100.00 &
    \incar{ICHARG=11}, \incar{NEDOS=2001}, \incar{LDAU=.TRUE.},
    \incar{LDAUL=2 0}, \incar{LDAUU=3.7 0} \\
    \bottomrule
  \end{tabularx}
\end{table}

These examples illustrate the intended division of work. Fine-tuning enables
the model to produce a task-relevant INCAR draft, while VASPGuard corrects
settings that can be decided by explicit workflow or materials rules. The
post-processor is not used to replace a missing calculation protocol.

\section{Code and data availability}

The accompanying release contains the complete training splits, the 4,682 source cases, benchmark, all training configurations, the
common local-generation and scoring code, the VASPGuard source, cleaning and audit reports, and a SHA-256 manifest. Base-model weights are obtained from the official Qwen, Llama, and Gemma distributions and remain subject to their respective licenses. LoRA adapters are separate weight artifacts and are not part of the source-code archive.

The complete release is available at \url{https://github.com/zxy-boop/VASPINCAR-Gen}. Code is released under the MIT License. The datasets---including the INCAR Training Set and the INCARBench evaluation split, both derived in part from Materials Project structures and reference INCAR files---are released under the Creative Commons Attribution 4.0 International License (CC BY 4.0), with attribution to the Materials Project as required by its terms of use. File integrity can be verified against the accompanying SHA-256 manifest.

\end{document}